# Wafer-Scale $MgB_2$ Superconducting Devices

**Changsub Kim[1]*, Christina Bell[1,2], Jake M. Evans[3], Jonathan Greenfield[1,4], Emma Batson[5], Karl K. Berggren[5], Nathan S. Lewis[3] & Daniel P. Cunnane[1]***

[1] Jet Propulsion Laboratory, California Institute of Technology, Pasadena, CA, USA
[2] Department of Physics, Arizona State University, Tempe, AZ, USA
[3] Division of Chemistry and Chemical Engineering, California Institute of Technology, Pasadena, CA, USA
[4] School of Earth and Space Exploration, Arizona State University, Tempe, AZ, USA
[5] Department of Electrical Engineering and Computer Science, Massachusetts Institute of Technology, Cambridge, MA, USA

**Progress in superconducting device and detector technologies over the past decade have realized practical applications in quantum computers, detectors for far-infrared telescopes, and optical communications. Superconducting thin film materials, however, have remained largely unchanged, with aluminum still being the material of choice for superconducting qubits, and niobium compounds for high frequency/high kinetic inductance devices. Magnesium diboride ($MgB_2$), known for its highest transition temperature ($T_c$ = 39 K) among metallic superconductors, is a viable material for elevated temperature and higher frequency superconducting devices moving towards THz frequencies. However, difficulty in synthesizing wafer-scale thin films have prevented implementation of $MgB_2$ devices into the application base of superconducting electronics. Here, we report ultra-smooth (< 0.5 nm root-mean-square roughness) and uniform $MgB_2$ thin (< 100 nm) films over 100 mm in diameter *for the first time* and present prototype devices fabricated with these films demonstrating key superconducting properties including internal quality factor over $10^4$ at 4.5 K and high tunable kinetic inductance in the order of tens of pH/sq in a 40 nm film. This groundbreaking advancement will enable development of elevated temperature, high frequency superconducting quantum circuits and devices.**

The quantum and nonlinear nature of superconductors has been of scientific interest since the discovery of superconductivity. Many applications of superconducting phenomena using thin film nano- and microdevices have shown unparalleled sensitivity for both power[1–4] and coherent detectors,[5,6] quantum-limited amplification,[7,8] and computation (quantum supremacy).[9] Current state-of-the-art superconducting devices are based on tried and tested aluminum or niobium thin films due to the ease of deposition (single element chemistry) and fabrication. More recently, research has taken advantage of more novel compounds and doped materials like TiN,[10] NbTiN,[11] Mn doped Al,[12] and granular Aluminum (gr-Al)[13] for high nonlinear kinetic inductance and to tune the critical temperature for pair-breaking applications. However, because of their low transition temperature ($T_c$, 1.20 K for Al and 9.26 K for Nb), the devices operate not only at low temperatures but at low frequencies (< 90 GHz for Al and < 700 GHz for Nb) from small superconducting gaps (superconducting energy gap $\Delta = 1.764\, k_B T_c$ according to the Bardeen–Cooper–Schrieffer (BCS) theory). **Using higher $T_c$ films can allow higher temperature operation**, **higher frequency operation**, or **a combination of the two to better suit operational needs and resilience against external factors and noise** (for example, decoherence in superconducting qubits). $MgB_2$ has the highest $T_c$ of 39 K among metallic superconductors,[14] and two superconducting gaps, with the interaction parameters between these gaps dependent on film quality and orientation.[15] The superconducting properties (density of states, penetration depth, etc) will fall somewhere between the BCS model predictions for the two independent gaps ($\Delta_\pi$ ~2.2 meV, $\Delta_\sigma$ ~7 meV)[16] enabling device operations above 1 THz.[17]

$MgB_2$ thin film thermodynamics,[18,19] deposition,[20–23] and fabrication[24] have been studied extensively since the

*e-mail: chang.sub.kim@jpl.nasa.gov; daniel.p.cunnane@jpl.nasa.gov

discovery of superconductivity in the compound, and some promising prototypes have been demonstrated[25–30] but practical applications have not caught on due to lack of scalability, poor reproducibility of the films, and fabrication immaturity of the material. The macroscopic film properties that would enable wider adoption of the material include large scale uniformity, and roughness below 1 nm root-mean-square (rms), while maintaining good superconducting properties (high $T_c$ and $J_c$). Further, deposition on silicon wafers would enable direct integration of the material into the state-of-the-art processes and technologies developed by the semiconductor industry, such as photolithography, plasma etch, and packaging, on a large scale. Wafer-scale deposition of $MgB_2$ by reactive evaporation had been reported back in 2006, but there are no follow-up studies since, likely due to its thickness (500 nm), roughness (1-5 nm rms), and lack of uniformity around the rotating axis at the center of the wafer.[31] Utilization of a commonly available PVD technique like sputtering provides good uniformity. While many groups have demonstrated the capability to sputter $MgB_2$, none have matured into successful technologies, likely due to difficulty in fabricating devices from these films or even difficulty in achieving reproducible films. Here, we report large-scale $MgB_2$ thin films on Si substrates with 100 mm diameter with $T_{c,0}$ over 32 K, roughness below 0.5 nm rms, and 1-σ wafer uniformity of 97.73 %. $T_{c,0}$ can be as high as 37 K, approaching bulk values, if we relax the expectations on the film roughness. We further developed standardized processes for $MgB_2$ nano- and microdevices fabrication and demonstrate superconducting resonators with $Q_i$ over $10^4$ at 4.5 K, $J_c$ of 10 MA/cm$^2$ at 4.2 K and kinetic inductance which can be tuned from moderate to high levels, meeting a strict criterion to realistically achieve mature fabrication capabilities.

## RESULTS and DISCUSSION

### $MgB_2$ Thin Film Fabrication, Properties and Characterization

The overall flow of our $MgB_2$ thin film fabrication process utilizing magnetron sputtering is illustrated in Figure 1a, and the resulting superconducting $MgB_2$ thin films on silicon and sapphire wafers are shown in Figure 1b and c, respectively. Magnetron sputtering is widely available, easily scalable, and produces uniform films, but *in-situ* sputtered $MgB_2$ films resulted in low $T_c$[22,32] from oxidation, small grain size, and/or off-stoichiometry due to high vapor pressure and low sticking coefficient of magnesium at elevated temperatures (> 200 °C), as well as contamination. Post-annealing of these films shows improvements in $T_c$ but at the cost of roughness (> 10 nm rms). Room temperature deposition results in uniform distribution of magnesium but requires a post-annealing process. Annealing magnesium-boron composite film in vacuum resulted in evaporation of magnesium, rough surface, and transition temperature around 6 K. Magnesium evaporation is prevented by capping the composite film with a thin (tens of nm) layer of high melting temperature material such as tantalum or boron. Tantalum does not react with magnesium or boron at typical annealing temperatures below 800 °C, but cracks above 700 °C and needs to be removed for easily measuring superconducting properties. Boron, a dielectric material with a high melting point of 2076 °C, serves as a better capping layer. Surface boron oxide has a low melting point of 450 °C and provides a crack-free, viscous capping layer.[33] There are two potential byproducts between $MgB_2$ and boron: $MgB_4$ and $MgB_7$. Both have slightly higher formation energies compared to $MgB_2$,[34] so their formation would be minimal, and would not affect the measurement of superconducting properties even if a thin layer of the byproducts is formed, because they are dielectric materials unlike metallic $MgB_2$. We have developed and demonstrated fabrication maturity in removing these capping layers for device development and optimization. In our work we tried many deposition conditions and, mostly through optimizing the roughness of the film, we chose a co-deposited Mg-B precursor film at room temperature. Sputtering of boron is very challenging, and the high melting point leads to a very low deposition rate. Given the propensity for oxidation, we optimized the boron sputtering for maximum rate, then tuned the magnesium sputtering conditions and annealing process to achieve optimized films. In order to achieve smooth films, the as-deposited, pre-annealed Mg-B composite film must be as smooth as possible. A small amount of RF substrate bias (15 W) during deposition reduced the surface roughness from 1.74 nm to 0.34~0.45 nm (rms). Increasing the bias beyond this point gave denser boron films, but etched the Mg or heated the surface too much to achieve controllable stoichiometry in a co-sputtered film. We saw indications that even this low substrate bias was enough energy to induce some reaction between Mg and B as we started to see transitions in as-sputtered films before any post-annealing step, though the $T_{c,0}$ was below liquid helium temperatures.

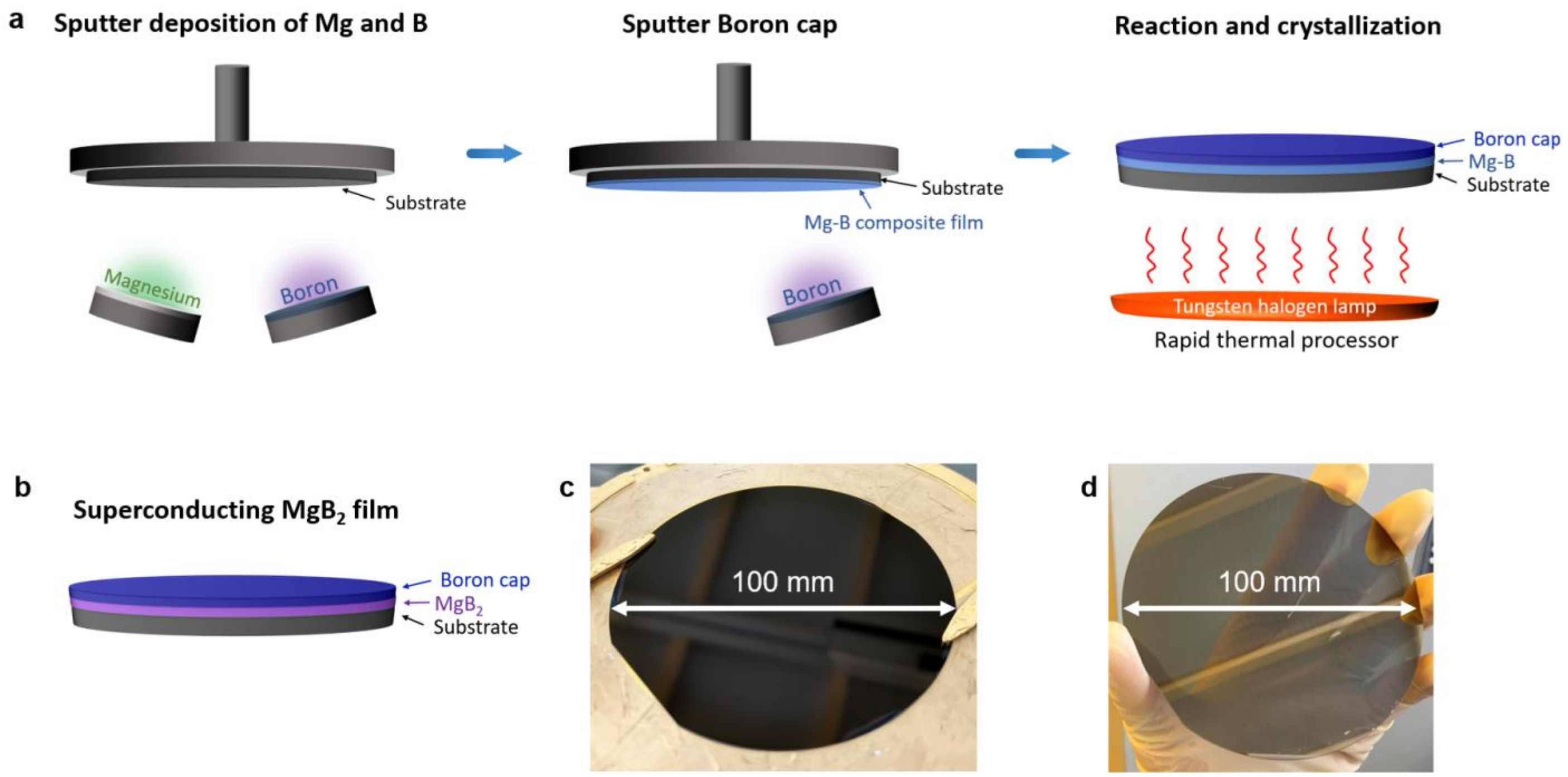

**Figure 1 | Schematic illustration of superconducting $MgB_2$ thin film fabrication process flow.** (a) Left: magnesium and boron are co-sputtered onto a rotating substrate to a desired thickness (e.g., 50 nm). A small substrate bias (e.g., 15 W) is applied to achieve smooth surface. Center: a thin boron capping layer is deposited on top of the co-sputtered film. Right: the wafer sample is annealed at around 600 °C in 100 % nitrogen environment for 2-10 minutes in a rapid thermal processor. (b) The final product is a superconducting thin $MgB_2$ film with a boron cap on a substrate. (c) $MgB_2$ thin film on a 100 mm diameter single crystal silicon substrate with a thin (30 nm) silicon nitride buffer layer. (d) $MgB_2$ thin film on a 100 mm diameter single crystal sapphire substrate.

A rapid thermal processor (RTP) provides good thermal uniformity across large wafers and was used for post-annealing. Because magnesium evaporation jumps up around its melting point of 650 °C, it is critical to keep the annealing temperature below this to avoid any potential surface roughening caused by evaporated magnesium that has yet to react with boron. By optimizing the annealing condition, we measured $T_{c,0}$ of 32 K (Figure 2a, b) for devices from $MgB_2$ thin films with 0.476 nm rms roughness (Figure 3a), and circular area of 100 mm diameter. We further achieved $T_{c,0}$ of 37.2 K, shown in Figure 2c and d, highest ever reported for sputtered film, by inducing large grains (thus sacrificing surface roughness) from post annealing interchanging magnesium and boron layers of stoichiometric ratio (2 boron to 1 magnesium atomic ratio), similar to previous attempts,[35–39] but with an additional boron capping layer on top. The massive migration of magnesium into the interfacing boron layers led to large $MgB_2$ grains and high $T_c$, but at the same time resulted in large voids and rough surfaces. Certain applications may take advantage of these films, particularly when devices have a very small active area.

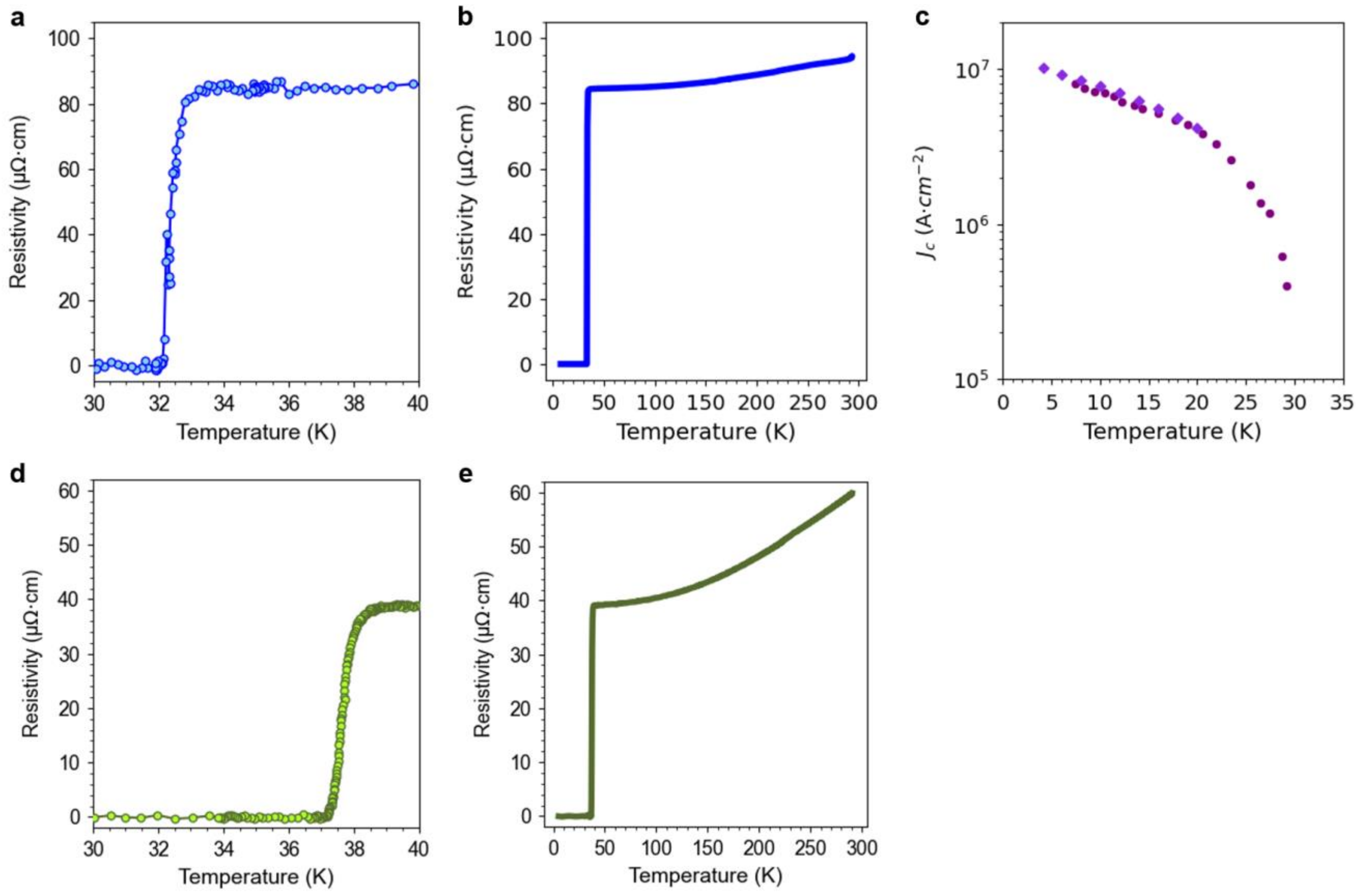

**Figure 2 | DC superconducting properties of $MgB_2$ thin films.** (a) Resistivity versus temperature plot of an $MgB_2$ thin film showing superconducting transition with $T_{c,0}$ = 32 K, (b) shown across a wider temperature range from 4.2 K to 300 K. (c) Critical current density ($J_c$) of two superconducting $MgB_2$ thin films at different temperatures ($J_c$ > $10^7$ MA·cm$^{-2}$ at 4.2 K) showing reproducibility of the films. (d) Resistivity versus temperature plot of $MgB_2$ film with $T_{c,0}$ = 37.2 K, highest ever reported for sputtered $MgB_2$ film, (e) shown across a wider temperature range from 4.2 K to 300 K.

Sapphire substrates have often been used for growing $MgB_2$ films because both have hexagonal crystal structure and lattice mismatch is less than 0.1 % with 30° rotation.[40] However, oxygen readily diffuses from sapphire to $MgB_2$ because magnesium has lower oxidation enthalpy compared to aluminum[41] and forms an interdiffusion layer.[17] This is especially detrimental for thin films under 100 nm. An alternative substrate is hexagonal SiC, especially for in-situ growth of $MgB_2$, because of the close lattice match. In fact, the highest $T_c$ ever measured for $MgB_2$ is from highly textured $MgB_2$ thin film slightly strained by SiC.[23] However, SiC is more expensive than Si, and polycrystalline $MgB_2$ films from post-annealing process would not be able to take advantage of the lattice match. Because magnesium reacts with silicon to form $Mg_2Si$, a thin inert buffer layer was used to prevent direct contact between $MgB_2$ and Si. As was demonstrated previously,[31] silicon nitride proved to be unreactive against magnesium, and LPCVD nitride as thin as 30 nm has been successfully tested to be sufficiently thick enough to serve as a good buffer, as shown in the STEM image in Figure 3b. In theory, the thickness is only limited to the threshold where no pinholes are left exposing Si, mostly dominated by the initial substrate roughness. The PECVD nitride deposition on Si wafers is a highly commercialized process,[42] not requiring the development of an independent recipe specific to $MgB_2$ thin films. Early in our work, we found that identical deposition and annealing recipes resulted in higher critical temperature on a silicon nitride buffer than on sapphire. Later it was confirmed through x-ray photoelectron spectroscopy (XPS) that magnesium from $MgB_2$ layer migrated to sapphire substrate and resulted in large deviation in stoichiometry of the film near film/substrate interface (Figure 3c).

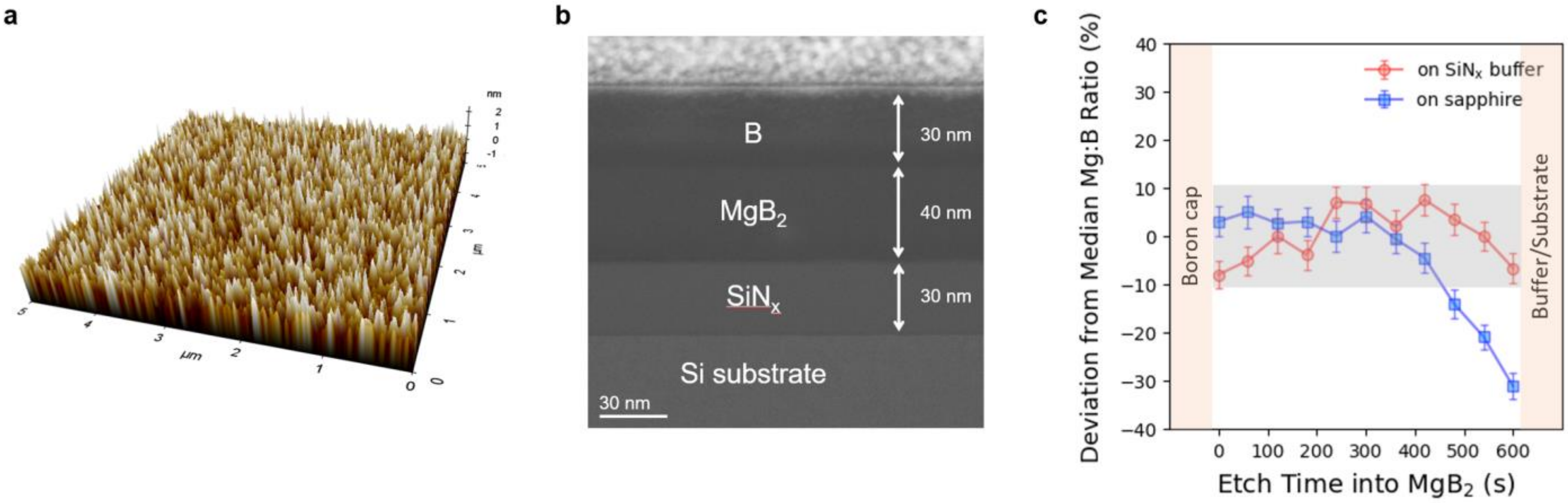

**Figure 3 | Morphological characterizations of post-annealed $MgB_2$ thin films.** (a) Atomic force microscopy of the $MgB_2$ thin film with $T_{c,0}$ of 32 K (Figure 2a) and the surface roughness of 0.476 nm rms. (b) High-angle annular dark-field (HAADF) STEM image of 40 nm thick superconducting $MgB_2$ thin film with 30 nm boron cap layer on high-resistivity silicon wafer with 30 nm silicon nitride buffer layer, showing sharp interfaces. (c) Deviation from median magnesium to boron ratio in 40 nm thick boron-rich $MgB_2$ film samples on silicon nitride buffer layer (red) and sapphire (blue) analyzed by depth-profile x-ray photoelectron spectroscopy. Magnesium to boron ratio stays within 10% of median for samples on silicon nitride. Significant migration of magnesium from $MgB_2$ layer to sapphire results in huge deviation of -30% from the median ratio near the interface. The range of deviation from median Mg:B ratio for the sample on silicon nitride is shaded in grey.

$MgB_2$ film uniformity is confirmed by mapping sheet resistance measurements by Eddy current on an insulating wafer, such as high-resistivity silicon or sapphire. As-deposited Mg-B composite films and post rapid-annealed $MgB_2$ thin films on 100mm wafers shown in Figure 4a and b demonstrate 1-σ wafer uniformities of 97.52% and 97.73% before and after annealing. The nonuniformity of as-deposited films is limited by the target size and the geometric configuration of the sputtering system and could be improved with larger targets. Post-annealed films' sheet resistance distribution change comes from the thermal gradient inside rapid thermal processors.

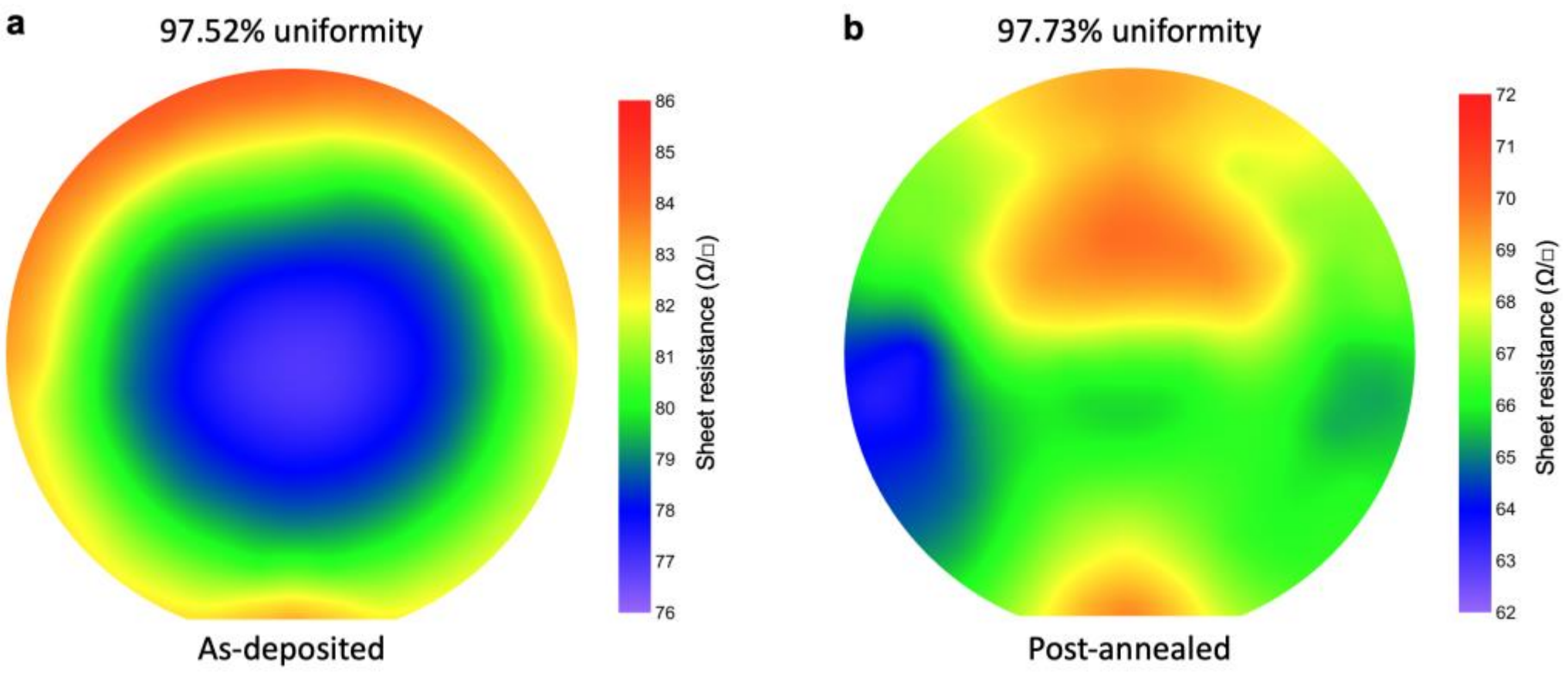

**Figure 4 | Eddy current maps of (a) as-deposited Mg-B composite film and (b) post-annealed superconducting $MgB_2$ film after annealing at 600 °C for 10 minutes.** Sheet resistances from eddy current measurements directly correspond to magnesium (as-deposited) and $MgB_2$ (post-annealed) distributions. The wafer uniformities (% of wafer within 1-σ of the average sheet resistance) are 97.52% and 97.73% before and after annealing, respectively.

For device design and fabrication, controlling resistivity and related materials properties such as kinetic inductance brings a huge advantage. Typically, such control is accomplished by deliberately adding impurities[43] or changing stoichiometry.[44] However, in materials deposited through reactive sputtering, such as NbN or TiN, the gas nonuniformity and hysteretic parameter curve make tuning unrealistic. The high $T_c$ $MgB_2$ thin films therefore have a tremendous advantage when it comes to the range of resistivity/kinetic inductance control.[30] Resistivity control in $MgB_2$ can be done by controlling the magnesium to boron ratios. Adding more boron or reducing the amount of magnesium by reducing the RF power on magnesium target, increases the resistivity of as-deposited/pre-annealed Mg-B composite films gradually, as shown in Figure 5. Annealing the films causes magnesium particles to react with boron and form $MgB_2$, resulting in a linear correlation between the resistivity of percolating magnesium (as-deposited) and percolating $MgB_2$ (post-annealed).

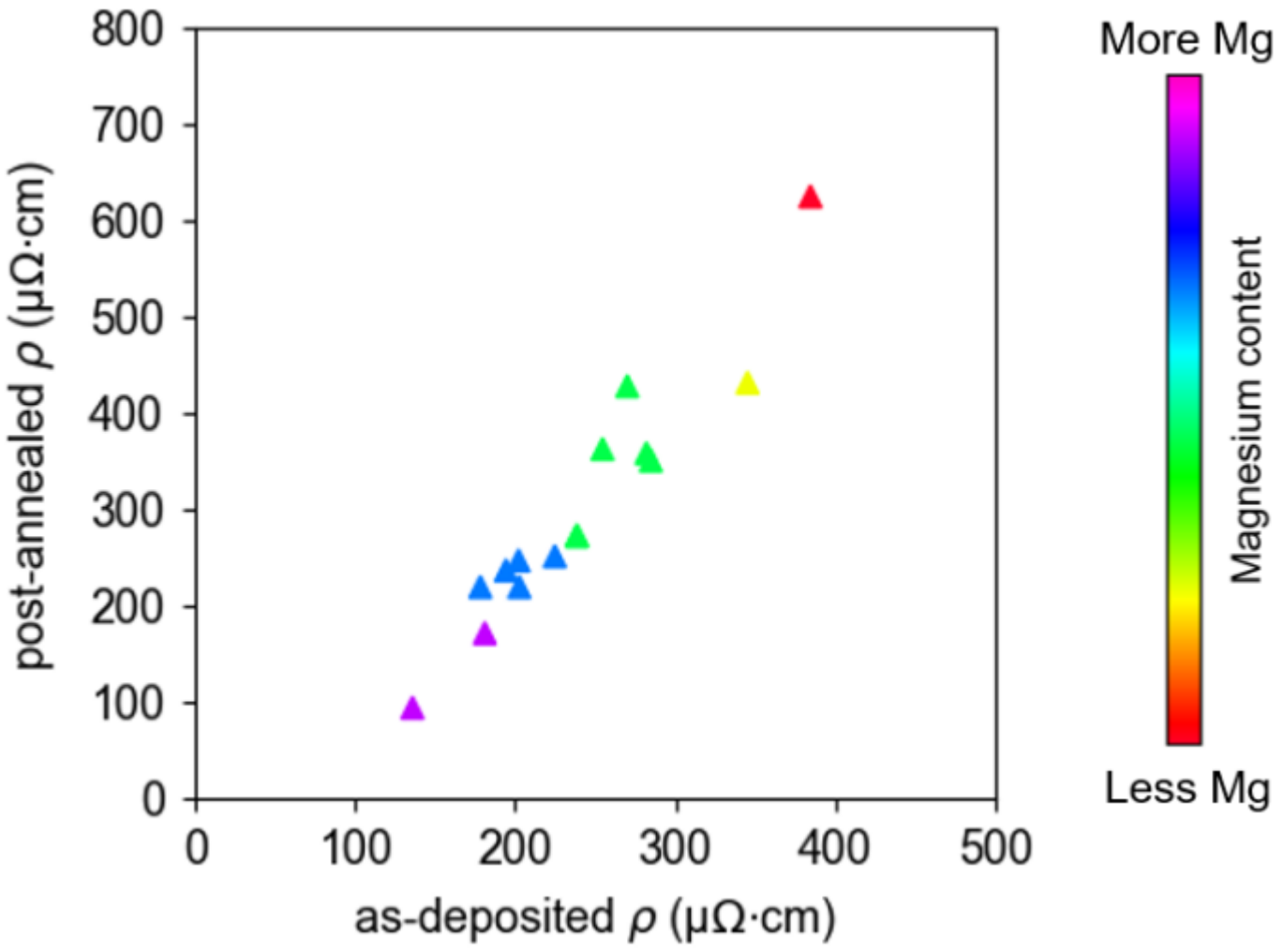

**Figure 5 | Post-annealed resistivity of smooth (roughness < 0.5 nm rms) $MgB_2$ films (y-axis) controlled by initial magnesium content in the as-deposited Mg-B composite films (resistivity along the x-axis).** Reducing the magnesium content increases as-deposited resistivity as well as post-annealed resistivity by forming boron-rich $MgB_2$ films.

## Superconducting Microdevices from Wafer-scale $MgB_2$ Films

Here we present and discuss results describing the internal quality factor and kinetic inductance of planar quarter-wave resonators patterned into our $MgB_2$ thin films using a coplanar waveguide (CPW) architecture. The overall flow of the device fabrication process is illustrated in Figure 6, fabricated wafer-scale devices are shown in Figure 7, and measurements and analysis of the devices in Figure 8. We demonstrate fabrication maturity on par with more commonly used materials (e.g. NbTiN)[45] and show figures of merit for microwave applications at higher temperatures. Given that the BCS surface resistance is both frequency and temperature dependent, better performance at lower frequencies and higher temperatures is a strong case for better performance for higher frequency at lower temperatures. Work is ongoing to improve on dielectric losses in the film stack, as well as demonstrate high frequency losses in the material and will be published following those results. The losses in a superconducting resonator can be measured through the quality factor of the resonator. By fitting the resonator to a well-known model,[46] we can obtain the internal microwave losses in the resonator. These losses include superconducting losses (from the BCS surface resistance), dielectric losses (both in the substrate/passivation layers, as well as the interfaces), and radiative losses. While our initial goal of this experiment was to try to measure the BCS surface resistance in the films, through troubleshooting and optimizing the losses, we have concluded that we are still dominated by dielectric losses in our thin film stack. By adopting standard fabrication techniques like a buffered oxide etch, we continue to see decreases in these losses, with the $Q_i \sim 10^4$ at 4.5 K (Figure 8d). We assume that the losses are limited by dielectrics or interfaces for two reasons. First, as we cool to lower temperature, the value of $Q_i$ has no temperature dependence from 1.5 K down to 1 K, with the value about $4\times10^4$. Work is ongoing to measure these films below 1 K, which will help separate dielectric losses (specifically TLS) from quasi-particle losses. Another strong indicator that the losses in our devices are indeed limited by dielectrics (or interfaces) is that the quality factor is highly dependent on the width of the devices. Smaller devices have higher dielectric fill factors and so the dielectric loss contribution goes up. The inset of Figure 8d shows the quality factor of 5 resonators (widths are 3, 4, 5, 6, and 10 um) of the same length (5 mm) scaling from ~7500 to 10,000. In the case that these losses are in the superconductor, there would be an $f^2$ dependence and given that the widest traces yield the highest frequency resonators, they would have the lowest quality factors.

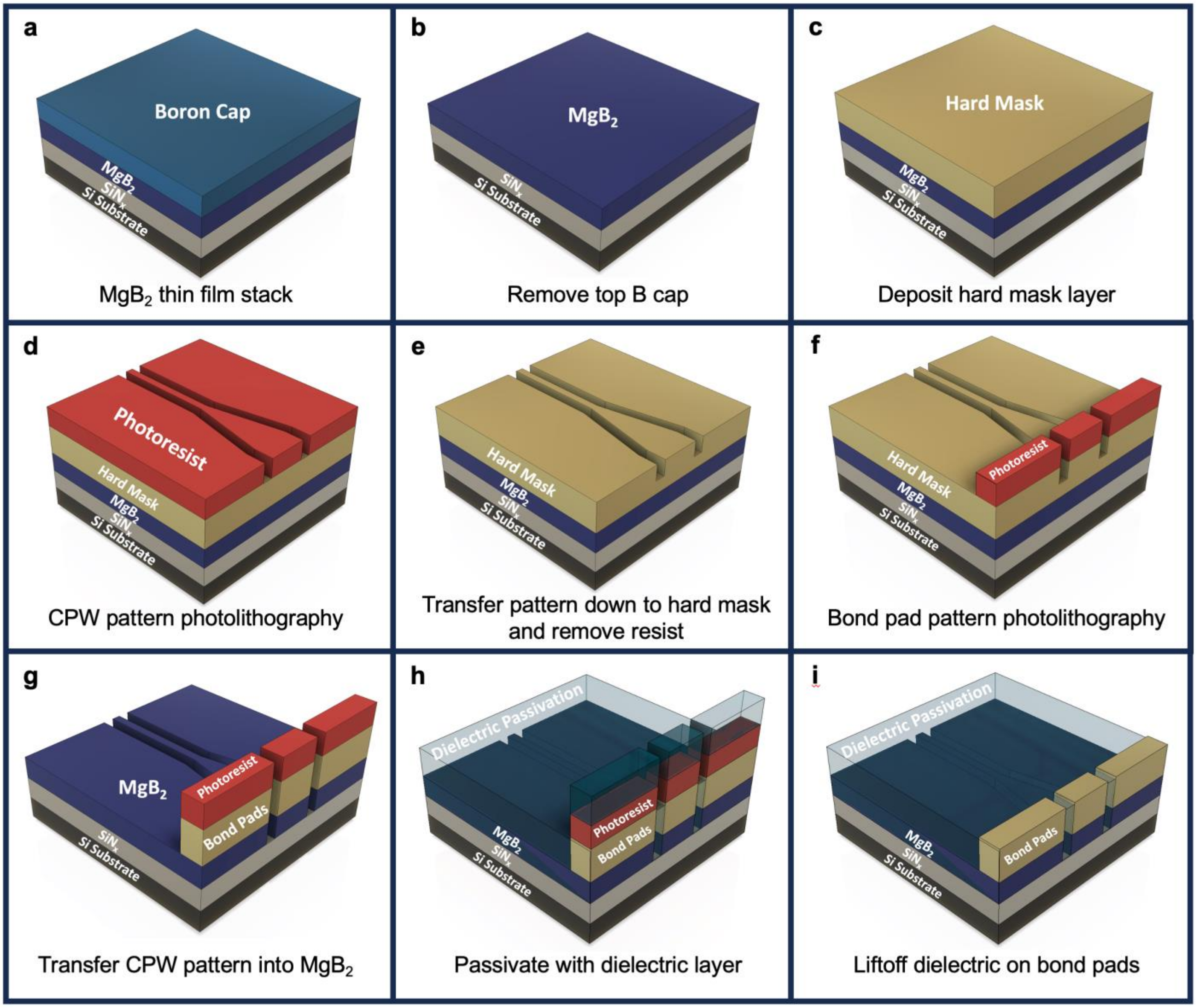

**Figure 6 | Schematic illustration of superconducting $MgB_2$ coplanar waveguide device fabrication process flow.**

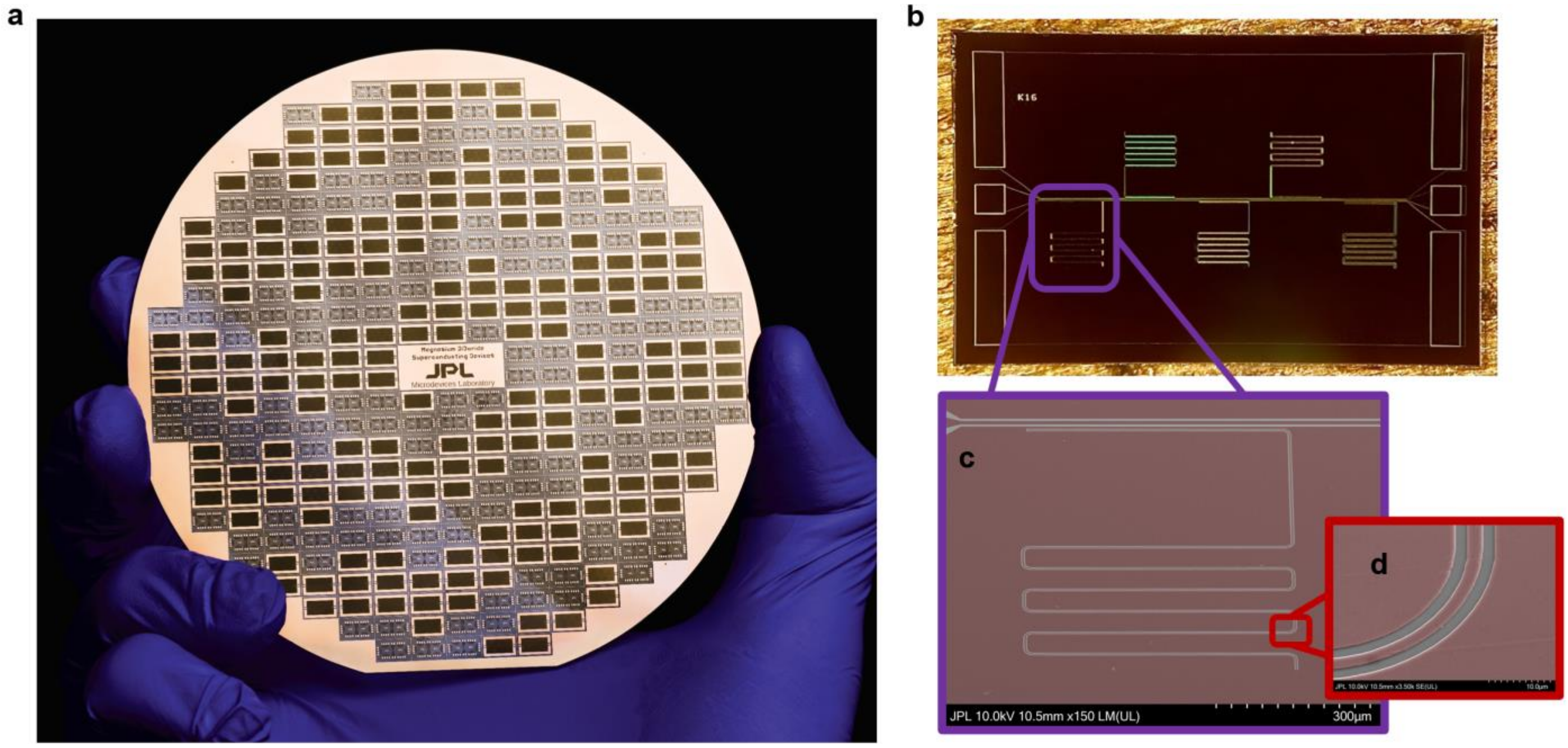

**Figure 7 | Wafer-scale $MgB_2$ superconducting device demonstration.** (a) Coplanar waveguide microwave Resonators and other test chips patterned in $MgB_2$ film deposited on 100mm sapphire substrate. (b) Optical image and (c, d) high magnification SEM images (false color) for a resonator chip on Si substrate. $MgB_2$ film is denoted by the red area.

These resonators also serve to demonstrate the high kinetic inductance ($L_k$) prevalent in these films. Since the geometric inductance and capacitance won't change much with different geometry of the same impedance, we can easily see how the kinetic inductance fraction α changes as a function of the width of the line. The design frequency of the resonator was simulated to be close to 7 GHz and the actual frequency of the resonances falls between 1.5 GHz and 2.8 GHz for conductor width from 3 µm to 10 µm. Comparing the design frequency with the actual frequency can give the value of alpha,[47] $\alpha = 1 - \left( f_{resonator} / f_{design} \right)^2$. Because $MgB_2$ doesn't have a typical temperature dependence, using the Mattis-Bardeen theory to estimate α is not feasible, but the method used is very accurate for the case of α > 0.5. In the same device used to describe the losses, alpha goes from 0.84 at 10 µm linewidth to 0.95 at 3 µm linewidth, with a consistent $L_k$ = 22~24 pH/□. This is shown in Figure 8b where two chips from different wafers are shown. In the second device, the same film recipe is used, but the films are slightly thinner and so $L_k$ = 28 pH/□. The latter shows even higher kinetic inductance can be readily achieved. By tuning the Mg content in the precursor film, we can achieve a range of 5-50 pH/□ for a similar film thickness (around 40 nm), since kinetic inductance is directly proportional to sheet resistance.[48] Using thinner films, we should be able to achieve very large numbers approaching or exceeding nH/□ values. A systematic study of RF properties as a function of Mg content is ongoing. In these films, tuning the stoichiometry is much more feasible than reactive gas sputtering, since we do not need to control gas partial pressures in a vacuum environment, which is a challenge in depositing Nb and Ti nitride films. We are also able to thin down the films using an argon ion mill or bromine plasma,[49] without degradation of the superconducting properties (within some limits). This has already been reported for producing even ultra-thin films of the same material.[50] Annealing a thicker film also avoids surface tension requiring more energy to achieve the same grain size and hence achieving the same critical temperature at lower temperatures.

We have used a fairly simple model that assumes a linear combination of the penetration depths associated with the two gaps in the material[29] to fit the temperature dependent frequency shift of two different resonators. The high coupling quality factor (~$10^4$) made it difficult to find the resonance above 10-15 K and so the data does not go beyond this temperature, however, the fits shown in Figure 8c show that the contribution from the larger gap is small (~12%), but non-negligible. This makes sense given that the smaller gap is responsible for the bulk of the kinetic inductance, which is high in our films. Using our thin film capability and expanding the two-gap model to include an interaction parameter which could also have temperature dependence, we hope to contribute to a global model for the contribution of the second gap to the overall film properties. Until now, film properties could not be reproduced under different conditions to modify the MB model for the two-gap material. Furthermore, there is only a rudimentary understanding on how these two gaps will affect high frequency performance in the films, and how pairbreaking of the smaller gap will contribute to RF losses while there still exists a larger gap. This is perhaps the most important future study planned for these films.

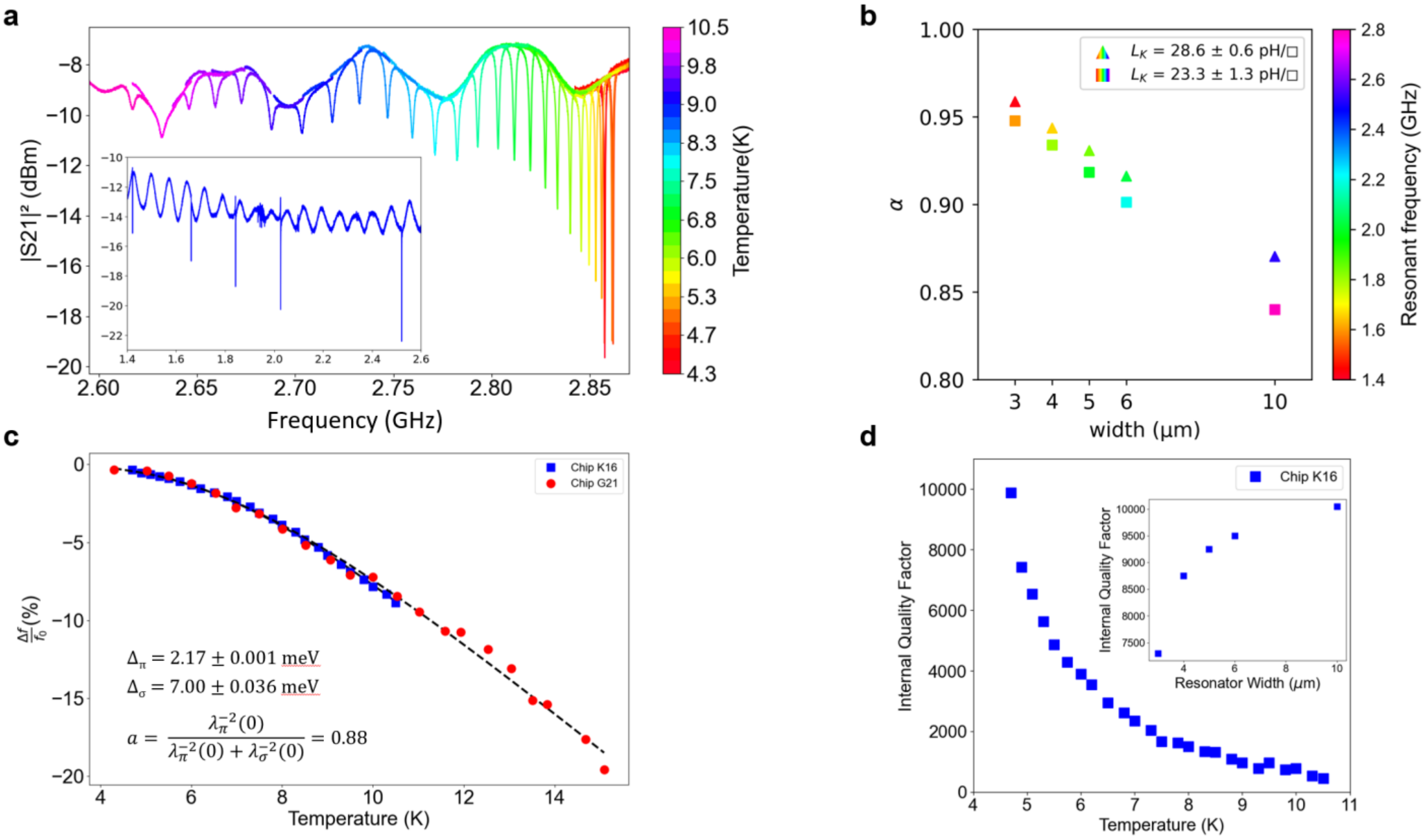

**Figure 8 | $MgB_2$ RF device measurements.** (a) S21 transmission of an $MgB_2$ resonator at different temperatures. Inset shows the five resonances corresponding to perfect yield on this chip. (b) Alpha (α) – fraction of total inductance that originates from the superconducting kinetic inductance – for two different chips (plotted in triangles and squares) at different CPW conductor widths. The resonant frequency for each measurement is represented by the color fill. (c) Fractional frequency shift as a function of temperature for two different resonator chips. Higher temperature data was not possible due to the high coupling factor to the resonators. In the future, we plan to design some low coupling resonators for better measurements up to the critical temperature. The fit is a crude 2-gap model used in Yang, et al.[29] The gaps values were highly constrained in the model to show the participation of the larger gap (about 12%) even for the polycrystalline films. (d) Internal quality factor ($Q_i$) as a function of temperature. We performed interface cleaning processes to achieve $Q_i \geq 10^4$ and it is expected to improve further as we mature the fabrication process. Inset shows the reduction in $Q_i$ seen for different cpw geometry (plotted as center conductor width for $Z_0$ = 50 Ohms).

## CONCLUSION

The ultra-smooth wafer-scale superconducting $MgB_2$ films developed here can be readily used for nano- and microdevice fabrication and by demonstrating key figures of merit – uniform and smooth large-scale film on Si substrate with high $T_c$, $J_c$, and $L_k$ – we have taken this technology to a level that can be immediately implemented. Historically, the low melting point and highly oxidizing nature of magnesium present multi-fold problems and have prevented any sustainable large-scale fabrication of $MgB_2$ devices. We made uniform and smooth large-area as-deposited Mg-B composite films by RF sputtering and substrate biasing. The films were grown on commercially available silicon substrate by adding a low-stress nitride buffer layer. By creating a closed environment with boron capping layer (which forms several nm of viscous crack-free liquid oxide cap) and prevented magnesium evaporation during annealing, we were able to optimize the annealing temperature and time by rapid thermal processor to achieve high $T_c$. We have further developed fabrication techniques for these films to achieve small critical features (0.5 μm demonstrated) and high yield. The devices demonstrate high $J_c$ and low rf losses suitable for high temperature/high frequency superconducting photon detectors and transmission lines. The large nonlinear kinetic inductance shown in these films opens a new branch of high frequency superconducting and quantum devices ranging from parametric amplifiers, frequency multipliers to kinetic inductance-based quantum computing qubits[51] and circuits.

## METHODS

Precursor Mg-B composite, layered thin films, and boron caps were deposited by RF magnetron sputtering of 3" elemental magnesium and boron targets at 1~2 mTorr and varying degrees of RF substrate biases. Magnesium sputtering power was varied between 50 W and 250 W RF for different recipes while boron sputtering power was fixed at 580 W RF. Silicon nitride buffer layer on high-resistivity silicon substrate was deposited by Tystar LPCVD with 50-150 MPa tensile stress. Post annealing was done using SSI Solaris 150 Rapid Thermal Processor with ramp rate of 1 degree per second for co-sputtered films and 1~33 degrees per second for layered films.

$MgB_2$ devices are fabricated by first removing the boron cap with argon ion mill. Because $MgB_2$ decomposes and oxidizes when exposed to an oxidant (e.g. air and water), passivation throughout device fabrication processes is necessary. We developed a tri-purpose titanium-gold bilayer, serving as an excellent passivation layer during fabrication (photolithography), a hardmask, and electrical contact pads after removal as a hardmask everywhere else on which a dielectric layer (silicon, silicon nitride, or aluminum nitride) is deposited for passivation. The devices are patterned by Heidelberg maskless aligner (MLA 150) with AZ5209-E or OiR 620 photoresists. After developing, the titanium layer not covered by photoresist is etched with chlorine ICP (10 sccm of $BCl_3$ and 35 sccm of $Cl_2$ at 7 mTorr with 200-400 W ICP power and 30 W substrate bias power). Once the pattern is transferred onto the titanium layer, photoresist is removed with acetone, then the entire sample is ion milled so that only $MgB_2$ device layer is left.

XPS depth profiling was done using Kratos Axis Ultra, with base pressure of $1\times10^{-9}$ Torr in the analysis chamber without etching or $5\times10^{-8}$ Torr during etching. A monochromatic Al $K_\alpha$ X-ray source with 1486.7 eV energy at 150 W was used. Etching was done with an $Ar^+$ ion beam with 2 kV beam energy over 2x2 $mm^2$ area. All spectra were collected with a resolution of 0.1 eV and pass energy of 20 eV and a 110 μm spot size during etching. Spectra were charge-corrected to adventitious C at 284.8 eV. Spectra were fit using CasaXPS software. All spectra used a Shirley background.

DC resistance and current density were measured using four-point method. RF measurements were done in liquid helium Dewar on a custom dipstick using a vector network analyzer.

We prepared the TEM lamella using focused ion beam milling (Raith Velion) with gold (Au+) from an $MgB_2$ sample that was deposited on a silicon wafer with a nitride buffer layer and post-annealed at 600 °C for 2 minutes. Platinum was deposited on top of the film as part of the lamella preparation. The lamella was heated in a vacuum oven at 150 °C overnight to minimize carbon contamination prior to imaging. The interfaces of the lamella were imaged in a 200 keV aberration-corrected STEM (Themis).

## ACKNOWLEDGEMENT

The research by C. Kim, C. Bell, J. Greenfield, and D. Cunnane. was carried out at the Jet Propulsion Laboratory, California Institute of Technology, under a contract with the National Aeronautics and Space Administration (80NM0018D0004). This work was primarily supported by Nancy Grace Roman Technology Fellowship in Astrophysics. We acknowledge the support and infrastructure provided for this work by the Microdevices Laboratory at JPL and The Kavli Nanoscience Institute at Caltech. We thank H. LeDuc, B. Bumble, and A. Beyer for advice on device fabrication, P. Day for discussions and support on RF measurements, M. Dickie for chemical vapor deposition of silicon nitride buffer layer, and A. Wertheim for discussions on sputter depositions. The XPS was carried out at the Molecular Materials Resource Center in the Beckman Institute at the California Institute of Technology and supported by U.S. Department of Energy grant numbers DE-SC0004993 and DE-SC0022087. The STEM work was completed in MIT.nano facilities. We thank Juan Ferrera for assistance in preparing the TEM lamella. E. Batson acknowledges the National Science Foundation Graduate Research Fellowship under Grant No. 2141064 and the NSF CQN program under Grant No. EEC1941583.

## REFERENCES

(1) Ade, P. A. R.; Ahmed, Z.; Amiri, M.; Barkats, D.; Thakur, R. B.; Bischoff, C. A.; Beck, D.; Bock, J. J.; Boenish, H.; Bullock, E.; Buza, V.; Cheshire IV, J. R.; Connors, J.; Cornelison, J.; Crumrine, M.; Cukierman, A.; Denison, E. V.; Dierickx, M.; Duband, L.; Eiben, M.; Fatigoni, S.; Filippini, J. P.; Fliescher, S.; Goeckner-Wald, N.; Goldfinger, D. C.; Grayson, J.; Grimes, P.; Hall, G.; Halal, G.; Halpern, M.; Hand, E.; Harrison, S.; Henderson, S.; Hildebrandt, S. R.; Hilton, G. C.; Hubmayr, J.; Hui, H.; Irwin, K. D.; Kang, J.; Karkare, K. S.; Karpel, E.; Kefeli, S.; Kernasovskiy, S. A.; Kovac, J. M.; Kuo, C. L.; Lau, K.; Leitch, E. M.; Lennox, A.; Megerian, K. G.; Minutolo, L.; Moncelsi, L.; Nakato, Y.; Namikawa, T.; Nguyen, H. T.; O'Brient, R.; Ogburn IV, R. W.; Palladino, S.; Prouve, T.; Pryke, C.; Racine, B.; Reintsema, C. D.; Richter, S.; Schillaci, A.; Schwarz, R.; Schmitt, B. L.; Sheehy, C. D.; Soliman, A.; Germaine, T. St.; Steinbach, B.; Sudiwala, R. V.; Teply, G. P.; Thompson, K. L.; Tolan, J. E.; Tucker, C.; Turner, A. D.; Umiltà, C.; Vergès, C.; Vieregg, A. G.; Wandui, A.; Weber, A. C.; Wiebe, D. V.; Willmert, J.; Wong, C. L.; Wu, W. L. K.; Yang, H.; Yoon, K. W.; Young, E.; Yu, C.; Zeng, L.; Zhang, C.; Zhang, S. BICEP/*Keck* XV: The BICEP3 Cosmic Microwave Background Polarimeter and the First Three-Year Data Set. *ApJ* **2022**, *927* (1), 77. https://doi.org/10.3847/1538-4357/ac4886.

(2) Day, P. K.; LeDuc, H. G.; Mazin, B. A.; Vayonakis, A.; Zmuidzinas, J. A Broadband Superconducting Detector Suitable for Use in Large Arrays. *Nature* **2003**, *425* (6960), 817–821. https://doi.org/10.1038/nature02037.

(3) Echternach, P. M.; Pepper, B. J.; Reck, T.; Bradford, C. M. Single Photon Detection of 1.5 THz Radiation with the Quantum Capacitance Detector. *Nat Astron* **2018**, *2* (1), 90–97. https://doi.org/10.1038/s41550-017-0294-y.

(4) Mattioli, F.; Zhou, Z.; Gaggero, A.; Gaudio, R.; Jahanmirinejad, S.; Sahin, D.; Marsili, F.; Leoni, R.; Fiore, A. Photon-Number-Resolving Superconducting Nanowire Detectors. *Supercond. Sci. Technol.* **2015**, *28* (10), 104001. https://doi.org/10.1088/0953-2048/28/10/104001.

(5) Zmuidzinas, J.; Ugras, N. G.; Miller, D.; Gaidis, M.; LeDuc, H. G.; Stern, J. A. Low-Noise Slot Antenna SIS Mixers. *IEEE Transactions on Applied Superconductivity* **1995**, *5* (2), 3053–3056. https://doi.org/10.1109/77.403236.

(6) Meledin, D.; Pavolotsky, A.; Desmaris, V.; Lapkin, I.; Risacher, C.; Perez, V.; Henke, D.; Nystrom, O.; Sundin, E.; Dochev, D.; Pantaleev, M.; Fredrixon, M.; Strandberg, M.; Voronov, B.; Goltsman, G.; Belitsky, V. A 1.3-THz Balanced Waveguide HEB Mixer for the APEX Telescope. *IEEE Transactions on Microwave Theory and Techniques* **2009**, *57* (1), 89–98. https://doi.org/10.1109/TMTT.2008.2008946.

(7) Macklin, C.; O'Brien, K.; Hover, D.; Schwartz, M. E.; Bolkhovsky, V.; Zhang, X.; Oliver, W. D.; Siddiqi, I. A near–Quantum-Limited Josephson Traveling-Wave Parametric Amplifier. *Science* **2015**, *350* (6258), 307–310. https://doi.org/10.1126/science.aaa8525.

(8) Eom, B. H.; Day, P. K.; LeDuc, H. G.; Zmuidzinas, J. A Wideband, Low-Noise Superconducting Amplifier with High Dynamic Range. *Nature Phys* **2012**, *8* (8), 623–627. https://doi.org/10.1038/nphys2356.

(9) Arute, F.; Arya, K.; Babbush, R.; Bacon, D.; Bardin, J. C.; Barends, R.; Biswas, R.; Boixo, S.; Brandao, F. G. S. L.; Buell, D. A.; Burkett, B.; Chen, Y.; Chen, Z.; Chiaro, B.; Collins, R.; Courtney, W.; Dunsworth, A.; Farhi, E.; Foxen, B.; Fowler, A.; Gidney, C.; Giustina, M.; Graff, R.; Guerin, K.; Habegger, S.; Harrigan, M. P.; Hartmann, M. J.; Ho, A.; Hoffmann, M.; Huang, T.; Humble, T. S.; Isakov, S. V.; Jeffrey, E.; Jiang, Z.; Kafri, D.; Kechedzhi, K.; Kelly, J.; Klimov, P. V.; Knysh, S.; Korotkov, A.; Kostritsa, F.; Landhuis, D.; Lindmark, M.; Lucero, E.; Lyakh, D.; Mandrà, S.; McClean, J. R.; McEwen, M.; Megrant, A.; Mi, X.; Michielsen, K.; Mohseni, M.; Mutus, J.; Naaman, O.; Neeley, M.; Neill, C.; Niu, M. Y.; Ostby, E.; Petukhov, A.; Platt, J. C.; Quintana, C.; Rieffel, E. G.; Roushan, P.; Rubin, N. C.; Sank, D.; Satzinger, K. J.; Smelyanskiy, V.; Sung, K. J.; Trevithick, M. D.; Vainsencher, A.; Villalonga, B.; White, T.; Yao, Z. J.; Yeh, P.; Zalcman, A.; Neven, H.; Martinis, J. M. Quantum Supremacy Using a Programmable Superconducting Processor. *Nature* **2019**, *574* (7779), 505–510. https://doi.org/10.1038/s41586-019-1666-5.

(10) Hubmayr, J.; Beall, J.; Becker, D.; Cho, H.-M.; Devlin, M.; Dober, B.; Groppi, C.; Hilton, G. C.; Irwin, K. D.; Li, D.; Mauskopf, P.; Pappas, D. P.; Van Lanen, J.; Vissers, M. R.; Wang, Y.; Wei, L. F.; Gao, J. Photon-Noise Limited Sensitivity in Titanium Nitride Kinetic Inductance Detectors. *Appl. Phys. Lett.* **2015**, *106* (7), 073505. https://doi.org/10.1063/1.4913418.

(11) Bretz-Sullivan, T. M.; Lewis, R. M.; Lima-Sharma, A. L.; Lidsky, D.; Smyth, C. M.; Harris, C. T.; Venuti, M.; Eley, S.; Lu, T.-M. High Kinetic Inductance NbTiN Superconducting Transmission Line Resonators in the Very Thin Film Limit. *Appl. Phys. Lett.* **2022**, *121* (5), 052602. https://doi.org/10.1063/5.0100961.

(12) Jones, G.; Johnson, B. R.; Abitbol, M. H.; Ade, P. A. R.; Bryan, S.; Cho, H.-M.; Day, P.; Flanigan, D.; Irwin, K. D.; Li, D.; Mauskopf, P.; McCarrick, H.; Miller, A.; Song, Y. R.; Tucker, C. High Quality Factor Manganese-Doped Aluminum Lumped-Element Kinetic Inductance Detectors Sensitive to Frequencies below 100 GHz. *Appl. Phys. Lett.* **2017**,

*110* (22), 222601. https://doi.org/10.1063/1.4984105.
(13) Grünhaupt, L.; Maleeva, N.; Skacel, S. T.; Calvo, M.; Levy-Bertrand, F.; Ustinov, A. V.; Rotzinger, H.; Monfardini, A.; Catelani, G.; Pop, I. M. Loss Mechanisms and Quasiparticle Dynamics in Superconducting Microwave Resonators Made of Thin-Film Granular Aluminum. *Phys. Rev. Lett.* **2018**, *121* (11), 117001. https://doi.org/10.1103/PhysRevLett.121.117001.
(14) Nagamatsu, J.; Nakagawa, N.; Muranaka, T.; Zenitani, Y.; Akimitsu, J. Superconductivity at 39 K in Magnesium Diboride. *Nature* **2001**, *410* (6824), 63–64. https://doi.org/10.1038/35065039.
(15) Brinkman, A.; Golubov, A. A.; Rogalla, H.; Dolgov, O. V.; Kortus, J.; Kong, Y.; Jepsen, O.; Andersen, O. K. Multiband Model for Tunneling in $MgB_2$ Junctions. *Phys. Rev. B* **2002**, *65* (18), 180517. https://doi.org/10.1103/PhysRevB.65.180517.
(16) Eskildsen, M. R.; Kugler, M.; Levy, G.; Tanaka, S.; Jun, J.; Kazakov, S. M.; Karpinski, J.; Fischer, Ø. Scanning Tunneling Spectroscopy on Single Crystal $MgB_2$. *Physica C: Superconductivity* **2003**, *385* (1–2), 169–176. https://doi.org/10.1016/S0921-4534(02)02301-8.
(17) Zeng, X.; Pogrebnyakov, A. V.; Kotcharov, A.; Jones, J. E.; Xi, X. X.; Lysczek, E. M.; Redwing, J. M.; Xu, S.; Li, Q.; Lettieri, J.; Schlom, D. G.; Tian, W.; Pan, X.; Liu, Z.-K. In Situ Epitaxial $MgB_2$ Thin Films for Superconducting Electronics. *Nature Mater* **2002**, *1* (1), 35–38. https://doi.org/10.1038/nmat703.
(18) Liu, Z.-K.; Schlom, D. G.; Li, Q.; Xi, X. X. Thermodynamics of the Mg–B System: Implications for the Deposition of $MgB_2$ Thin Films. *Appl. Phys. Lett.* **2001**, *78* (23), 3678–3680. https://doi.org/10.1063/1.1376145.
(19) Kim, J.; Singh, R. K.; Newman, N.; Rowell, J. M. Thermochemistry of $MgB_2$ Thin Film Synthesis. *IEEE Trans. Appl. Supercond.* **2003**, *13* (2), 3238–3241. https://doi.org/10.1109/TASC.2003.812210.
(20) Kang, W. N. $MgB_2$ Superconducting Thin Films with a Transition Temperature of 39 Kelvin. *Science* **2001**, *292* (5521), 1521–1523. https://doi.org/10.1126/science.1060822.
(21) Saito, A.; Kawakami, A.; Shimakage, H.; Wang, Z. As-Grown Deposition of Superconducting $MgB_2$ Thin Films by Multiple-Target Sputtering System. *Jpn. J. Appl. Phys.* **2002**, *41* (Part 2, No. 2A), L127–L129. https://doi.org/10.1143/JJAP.41.L127.
(22) Schneider, R.; Geerk, J.; Linker, G.; Ratzel, F.; Zaitsev, A. G.; Obst, B. In Situ Deposition of $MgB_2$ Thin Films by Magnetron Cosputtering and Sputtering Combined with Thermal Evaporation. *Physica C: Superconductivity and its Applications* **2005**, *423* (3–4), 89–95. https://doi.org/10.1016/j.physc.2005.04.004.
(23) Xi, X. X.; Pogrebnyakov, A. V.; Xu, S. Y.; Chen, K.; Cui, Y.; Maertz, E. C.; Zhuang, C. G.; Li, Q.; Lamborn, D. R.; Redwing, J. M.; Liu, Z. K.; Soukiassian, A.; Schlom, D. G.; Weng, X. J.; Dickey, E. C.; Chen, Y. B.; Tian, W.; Pan, X. Q.; Cybart, S. A.; Dynes, R. C. $MgB_2$ Thin Films by Hybrid Physical–Chemical Vapor Deposition. *Physica C: Superconductivity* **2007**, *456* (1–2), 22–37. https://doi.org/10.1016/j.physc.2007.01.029.
(24) Naito, M.; Ueda, K. $MgB_2$ Thin Films for Superconducting Electronics. *Supercond. Sci. Technol.* **2004**, *17* (7), R1–R18. https://doi.org/10.1088/0953-2048/17/7/R01.
(25) Mijatovic, D.; Brinkman, A.; Oomen, I.; Rijnders, G.; Hilgenkamp, H.; Rogalla, H.; Blank, D. H. A. Magnesium-Diboride Ramp-Type Josephson Junctions. *Appl. Phys. Lett.* **2002**, *80* (12), 2141–2143. https://doi.org/10.1063/1.1462869.
(26) Brinkman, A.; Rowell, J. M. $MgB_2$ Tunnel Junctions and SQUIDs. *Physica C: Superconductivity* **2007**, *456* (1), 188–195. https://doi.org/10.1016/j.physc.2007.01.019.
(27) Yoshioka, N.; Yagi, I.; Shishido, H.; Yotsuya, T.; Miyajima, S.; Fujimaki, A.; Miki, S.; Zhen Wang; Ishida, T. Current-Biased Kinetic Inductance Detector Using $MgB_2$ Nanowires for Detecting Neutrons. *IEEE Trans. Appl. Supercond.* **2013**, *23* (3), 2400604–2400604. https://doi.org/10.1109/TASC.2013.2243812.
(28) Cunnane, D.; Kawamura, J. H.; Acharya, N.; Wolak, M. A.; Xi, X. X.; Karasik, B. S. Low-Noise THz $MgB_2$ Josephson Mixer. *Appl. Phys. Lett.* **2016**, *109* (11), 112602. https://doi.org/10.1063/1.4962634.
(29) Yang, C.; Niu, R. R.; Guo, Z. S.; Cai, X. W.; Chu, H. M.; Yang, K.; Wang, Y.; Feng, Q. R.; Gan, Z. Z. Lumped Element Kinetic Inductance Detectors Based on Two-Gap $MgB_2$ Thin Films. *Appl. Phys. Lett.* **2018**, *112* (2), 022601. https://doi.org/10.1063/1.5013286.
(30) Cherednichenko, S.; Acharya, N.; Novoselov, E.; Drakinskiy, V. Low Kinetic Inductance Superconducting $MgB_2$ Nanowires with a 130 Ps Relaxation Time for Single-Photon Detection Applications. *Supercond. Sci. Technol.* **2021**, *34* (4), 044001. https://doi.org/10.1088/1361-6668/abdeda.
(31) Moeckly, B. H.; Ruby, W. S. Growth of High-Quality Large-Area $MgB_2$ Thin Films by Reactive Evaporation. *Supercond. Sci. Technol.* **2006**, *19* (6), L21–L24. https://doi.org/10.1088/0953-2048/19/6/L02.
(32) Shimakage, H.; Miki, S.; Tsujimoto, K.; Wang, Z.; Ishida, T.; Tonouchi, M. Characteristics of As-Grown $MgB_2$ Thin Films Made by Sputtering. *IEEE Trans. Appl. Supercond.* **2005**, *15* (2), 3269–3272. https://doi.org/10.1109/TASC.2005.848849.
(33) Napolitano, A.; Macedo, P. B.; Hawkins, E. G. Viscosity and Density of Boron Trioxide. *Journal of the American Ceramic Society* **1965**, *48* (12), 613–616. https://doi.org/10.1111/j.1151-2916.1965.tb14690.x.
(34) Balducci, G.; Brutti, S.; Ciccioli, A.; Gigli, G.; Manfrinetti, P.; Palenzona, A.; Butman, M. F.; Kudin, L. Thermodynamics of the Intermediate Phases in the Mg–B System. *Journal of Physics and Chemistry*

of Solids 2005, 66 (2–4), 292–297. https://doi.org/10.1016/j.jpcs.2004.06.063.
(35) Zhu, H. M.; Zhang, Y. B.; Sun, X. L.; Xiong, W. J.; Zhou, S. P. $MgB_2$ Thin Films on Si(111) without a Buffer Layer Prepared by e-Beam Evaporation. *Physica C: Superconductivity* **2007**, *452* (1–2), 11–15. https://doi.org/10.1016/j.physc.2006.11.009.
(36) Dai, Q.; Kong, X.; Feng, Q.; Yang, Q.; Zhang, H.; Nie, R.; Han, L.; Ma, Y.; Wang, F. $MgB_2$ Films Prepared by Rapid Annealing Method. *Physica C: Superconductivity* **2012**, *475*, 24–27. https://doi.org/10.1016/j.physc.2012.01.014.
(37) Chromik, Š.; Nishida, A.; Štrbík, V.; Gregor, M.; Espinós, J. P.; Liday, J.; Durný, R. The Distribution of Elements in Sequentially Prepared $MgB_2$ on SiC Buffered Si Substrate and Possible Pinning Mechanisms. *Applied Surface Science* **2013**, *269*, 29–32. https://doi.org/10.1016/j.apsusc.2012.10.019.
(38) Altin, E.; Kurt, F.; Altin, S.; Yakinci, M. E.; Yakinci, Z. D. $MgB_2$ Thin Film Fabrication with Excess Mg by Sequential e-Beam Evaporation and Transport Properties under Magnetic Fields. *Current Applied Physics* **2014**, *14* (3), 245–250. https://doi.org/10.1016/j.cap.2013.11.023.
(39) Beckham, J. L.; Bayu Aji, L. B.; Baker, A. A.; Bae, J. H.; Stavrou, E.; Jacob, R. E.; McCall, S. K.; Kucheyev, S. O. Superconducting Films of $MgB_2$ via Ion Beam Mixing of Mg/B Multilayers. *J. Phys. D: Appl. Phys.* **2020**, *53* (20), 205302. https://doi.org/10.1088/1361-6463/ab7624.
(40) Seong, W. K.; Oh, S.; Kang, W. N. Perfect Domain-Lattice Matching between $MgB_2$ and $Al_2O_3$: Single-Crystal $MgB_2$ Thin Films Grown on Sapphire. *Jpn. J. Appl. Phys.* **2012**, *51*, 083101. https://doi.org/10.1143/JJAP.51.083101.
(41) Hasegawa, M. Chapter 3.3 - Ellingham Diagram. In *Treatise on Process Metallurgy*; Seetharaman, S., Ed.; Elsevier: Boston, 2014; pp 507–516. https://doi.org/10.1016/B978-0-08-096986-2.00032-1.
(42) Mackenzie, K.; Johnson, D.; DeVre, M.; Westerman, R.; Reelfs, B. Stress Control of Si-Based PECVD Dielectrics. *Meet. Abstr.* **2006**, *MA2005-01* (9), 406–406. https://doi.org/10.1149/MA2005-01/9/406.
(43) Ruggiero, S. T.; Williams, A.; Rippard, W. H.; Clark, A. M.; Deiker, S. W.; Young, B. A.; Vale, L. R.; Ullom, J. N. Dilute Al–Mn Alloys for Superconductor Device Applications. *Nuclear Instruments and Methods in Physics Research Section A: Accelerators, Spectrometers, Detectors and Associated Equipment* **2004**, *520* (1), 274–276. https://doi.org/10.1016/j.nima.2003.11.236.
(44) Yemane, Y. T.; Sowa, M. J.; Zhang, J.; Ju, L.; Deguns, E. W.; Strandwitz, N. C.; Prinz, F. B.; Provine, J. Superconducting Niobium Titanium Nitride Thin Films Deposited by Plasma-Enhanced Atomic Layer Deposition. *Supercond. Sci. Technol.* **2017**, *30* (9), 095010. https://doi.org/10.1088/1361-6668/aa7ce3.
(45) Malnou, M.; Vissers, M. R.; Wheeler, J. D.; Aumentado, J.; Hubmayr, J.; Ullom, J. N.; Gao, J. Three-Wave Mixing Kinetic Inductance Traveling-Wave Amplifier with Near-Quantum-Limited Noise Performance. *PRX Quantum* **2021**, *2* (1), 010302. https://doi.org/10.1103/PRXQuantum.2.010302.
(46) Carter, F. W.; Khaire, T. S.; Novosad, V.; Chang, C. L. Scraps: An Open-Source Python-Based Analysis Package for Analyzing and Plotting Superconducting Resonator Data. *IEEE Transactions on Applied Superconductivity* **2017**, *27* (4), 1–5. https://doi.org/10.1109/TASC.2016.2625767.
(47) Gao, J. The Physics of Superconducting Microwave Resonators, California Institute of Technology, Pasadena, CA, 2008.
(48) Annunziata, A. J.; Santavicca, D. F.; Frunzio, L.; Catelani, G.; Rooks, M. J.; Frydman, A.; Prober, D. E. Tunable Superconducting Nanoinductors. *Nanotechnology* **2010**, *21* (44), 445202. https://doi.org/10.1088/0957-4484/21/44/445202.
(49) Shibata, H. Fabrication of a $MgB_2$ Nanowire Single-Photon Detector Using $Br_2$–$N_2$ Dry Etching. *Appl. Phys. Express* **2014**, *7* (10), 103101. https://doi.org/10.7567/APEX.7.103101.
(50) Acharya, N.; Wolak, M. A.; Tan, T.; Lee, N.; Lang, A. C.; Taheri, M.; Cunnane, D.; Karasik, Boris. S.; Xi, X. X. $MgB_2$ Ultrathin Films Fabricated by Hybrid Physical Chemical Vapor Deposition and Ion Milling. *APL Materials* **2016**, *4* (8), 086114. https://doi.org/10.1063/1.4961635.
(51) Faramarzi, F.; Day, P.; Glasby, J.; Sypkens, S.; Colangelo, M.; Chamberlin, R.; Mirhosseini, M.; Schmidt, K.; Berggren, K. K.; Mauskopf, P. Initial Design of a W-Band Superconducting Kinetic Inductance Qubit. *IEEE Trans. Appl. Supercond.* **2021**, *31* (5), 1–5. https://doi.org/10.1109/TASC.2021.3065304.

## Corresponding Authors

*chang.sub.kim@jpl.nasa.gov
*daniel.p.cunnane@jpl.nasa.gov

## Author contributions

C.K. and D.P.C. conceived and designed the experimental protocol. C.K. prepared and optimized the films. C.K. performed eddy current, cryogenic DC conductivity, and AFM measurements. D.P.C. designed the resonator patterns. C.K. fabricated the devices. C.B., J.G. and D.P.C. performed cryogenic RF measurements. C.K. and D.P.C. interpreted the results. J.M.E. performed XPS depth-profiling and analyzed the data. N.S.L. secured funding for the XPS equipment and analysis. E.B. performed sampling and STEM imaging of the $MgB_2$ lamella. K.K.B. supervised E.B.'s work. D.P.C. provided guidance throughout the project. C.K. and D.P.C. wrote the manuscript. All the co-authors discussed the results and helped revise the manuscript.

**Conflict of Interest Disclosure**

The California Institute of Technology has filed a U.S. utility patent with the title "Wafer scale production of superconducting magnesium diboride thin films with high transition temperature" (inventors: C.K. and D.P.C.) describing the superconducting magnesium diboride thin film and device fabrication methods described in this paper.

Supporting Information

# Wafer-Scale $MgB_2$ Superconducting Devices

**Changsub Kim[1*], Christina Bell[1,2], Jake M. Evans[3], Jonathan Greenfield[1,4], Emma Batson[5], Karl K. Berggren[5], Nathan S. Lewis[3] & Daniel P. Cunnane[1*]**

[1] Jet Propulsion Laboratory, California Institute of Technology, Pasadena, CA, USA
[2] Department of Physics, Arizona State University, Tempe, AZ, USA
[3] Division of Chemistry and Chemical Engineering, California Institute of Technology, Pasadena, CA, USA
[4] School of Earth and Space Exploration, Arizona State University, Tempe, AZ, USA

X-ray photoelectron spectroscopy (XPS) depth profile of $MgB_2$ thin films used to create Figure 3c

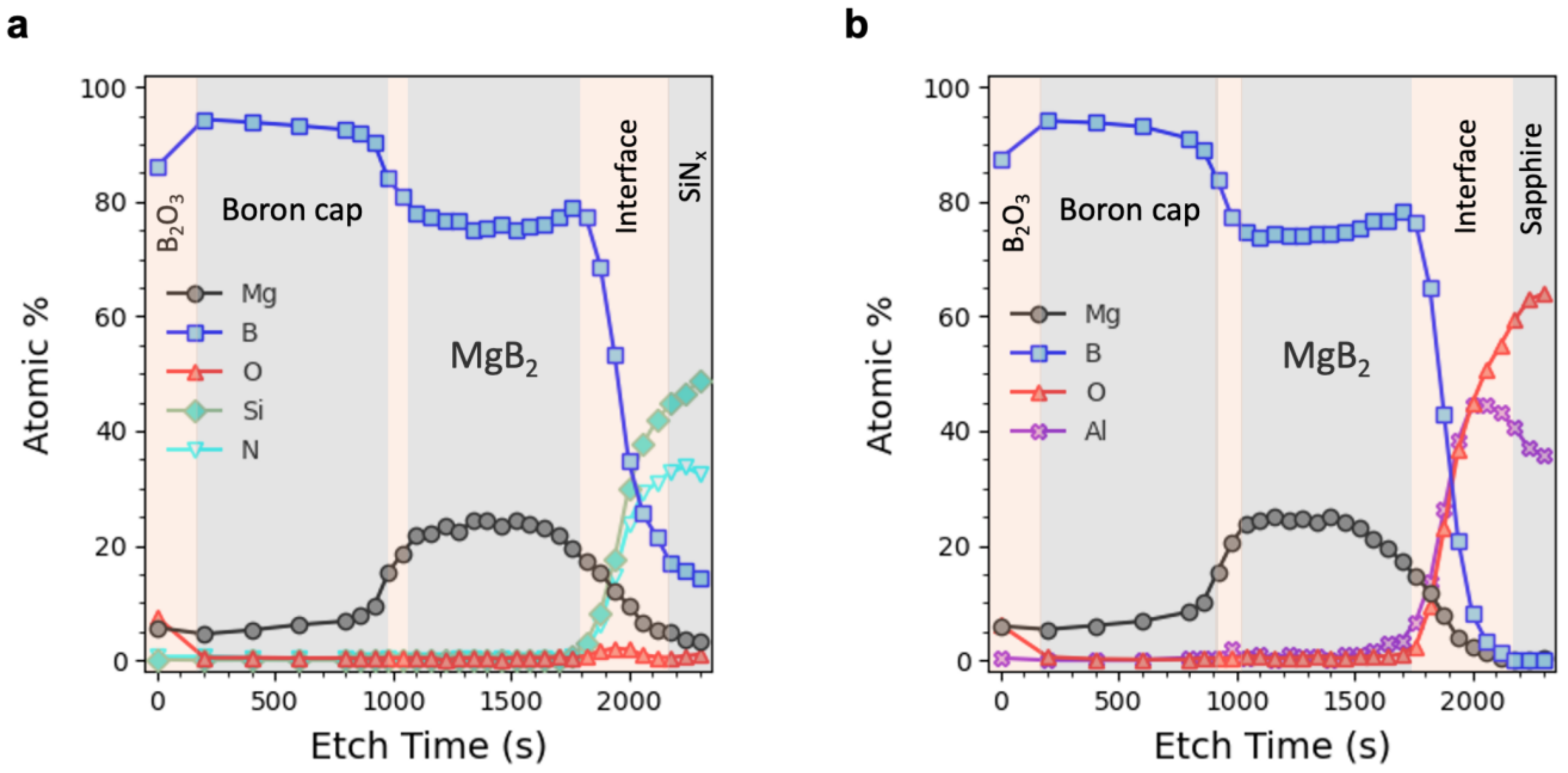

**Figure S1 | Atomic percentages of different elements in $MgB_2$ film sample stack on (a) silicon nitride buffer on silicon, and (b) sapphire by depth profile XPS. Both samples have boron caps with surface oxide ($B_2O_3$) and few atomic % of Mg, indicating there is some diffusion of magnesium into the capping layer, likely forming $MgB_4$ or $MgB_7$. The atomic % of Mg in $MgB_2$ layer is generally flat in (a) considering depth resolution of XPS, but clearly shows a decrease towards $MgB_2$/sapphire interface in (b). In (a), there is some diffusion of boron from $MgB_2$ layer to silicon nitride layer, but no diffusion of silicon or nitrogen from the nitride layer to the $MgB_2$ layer. Etch times do not correspond directly to layer thicknesses, since etch rates are different for each layer, with $MgB_2$ being the highest, followed by boron, then sapphire and silicon nitride.**

*e-mail: chang.sub.kim@jpl.nasa.gov; daniel.p.cunnane@jpl.nasa.gov

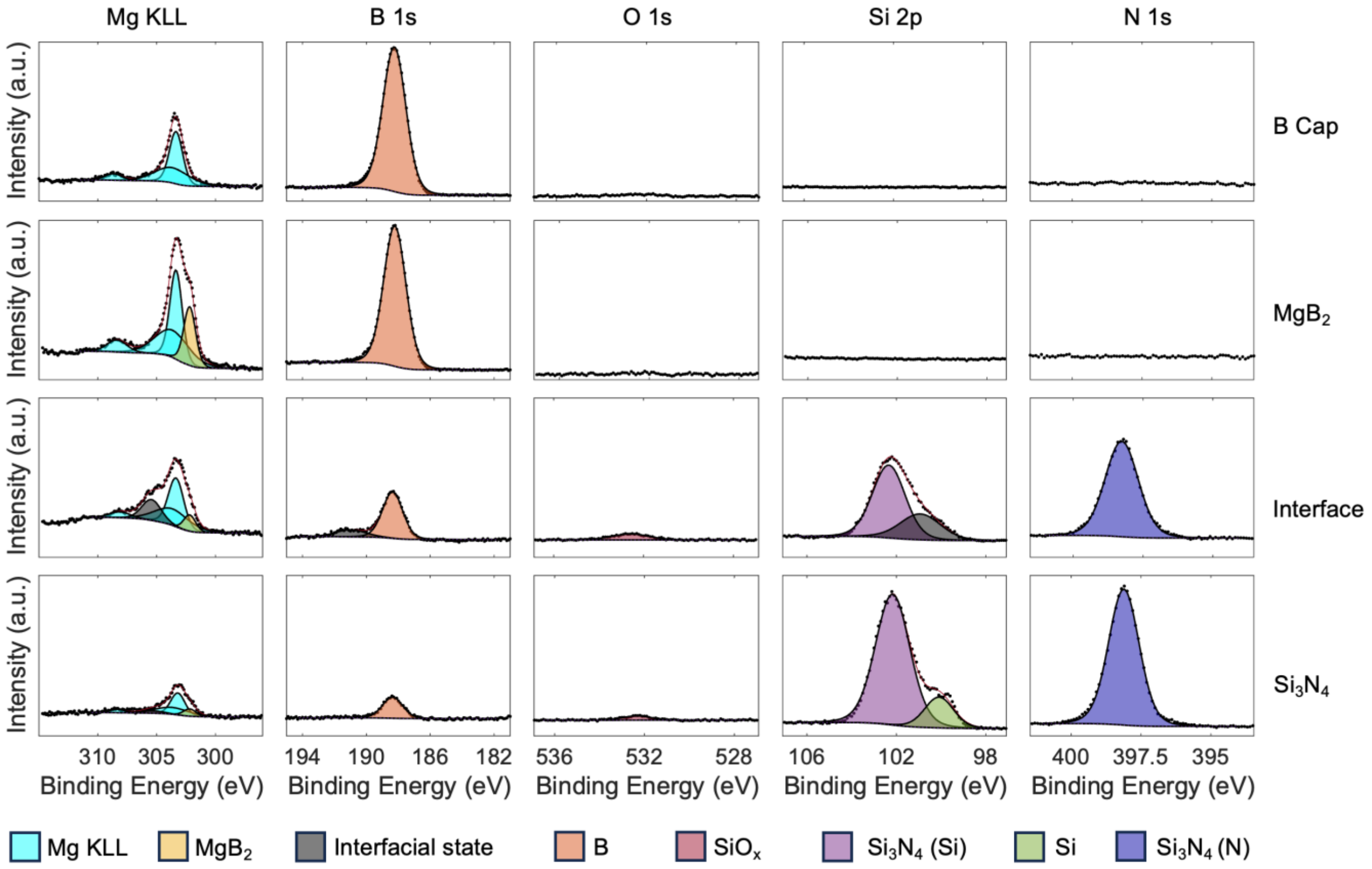

**Figure S2 | Representative x-ray photoelectron (XP) spectra (columns) of Mg KLL, B 1s, O 1s, Si 2p and N 1s in each layer of $MgB_2$ sample stack on silicon nitride as denoted by row labels. Mg KLL transitions are not assigned here due to uncertainty in the composition and intensity of transitions in the observed chemistry. In the B cap layer, Mg KLL displayed 3 peaks, but likely originating from a Mg boride ($MgB_2$, $MgB_4$ or $MgB_7$) as no other elements were detected in this region. In the $MgB_2$ layer, an additional peak is found at a lower BE (higher KE, in Auger convention) which is likely from $MgB_2$. At the $MgB_2/Si_3N_4$ interface, an additional Mg KLL peak is observed at ~306 eV BE along with a B 1s peak at ~191 eV and a Si 2p peak at ~101 eV as an O impurity is detected. Each of these observations is consistent with an oxide impurity, likely consisting of Mg, B, and Si, but not N. Once the interface is etched through, a small amount of Mg, B, and O is still detected, but the primary signals are $Si_3N_4$ and underlying Si.**

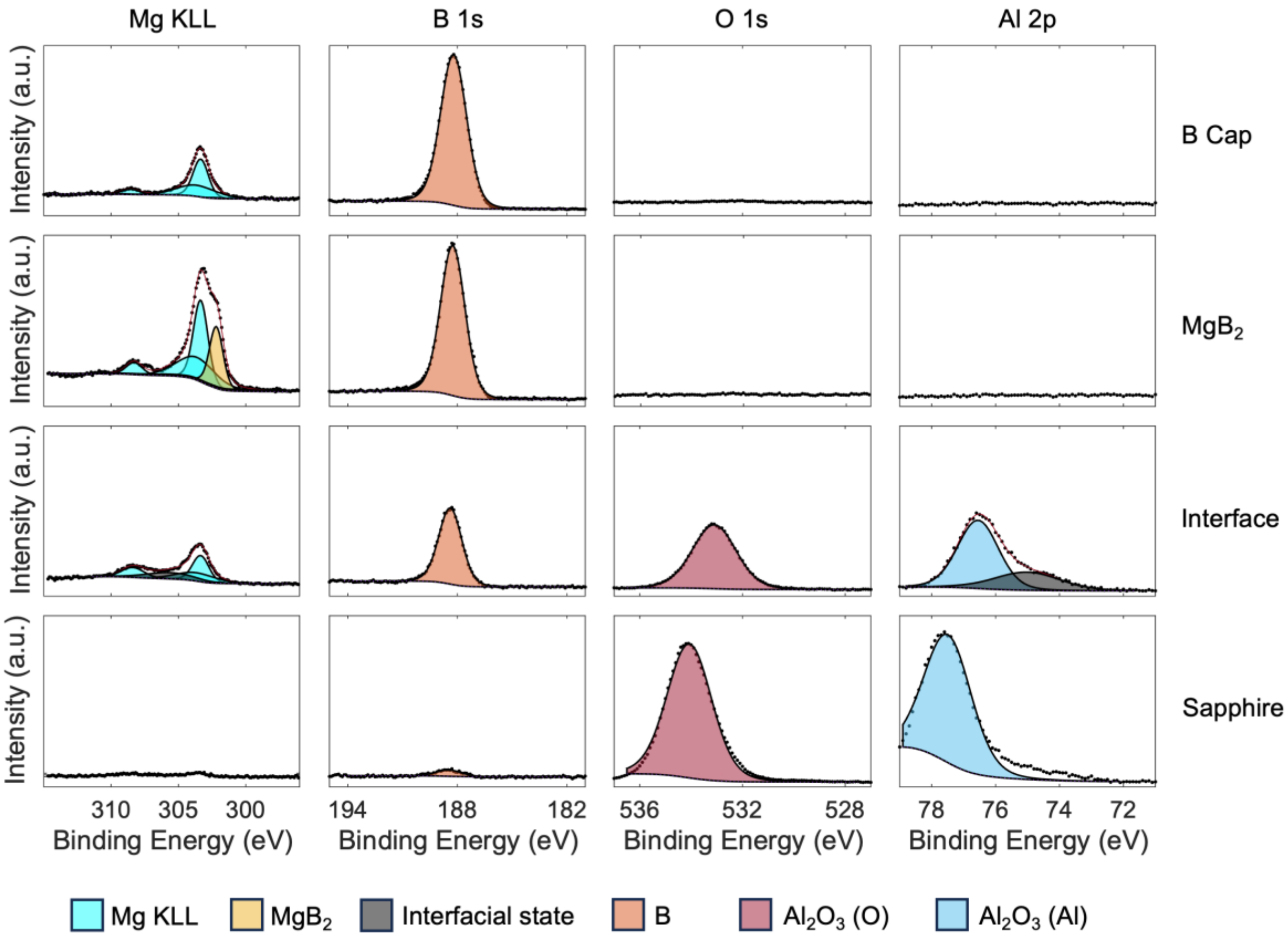

**Figure S3| Representative XP spectra (columns) of Mg KLL, B 1s, O 1s, and Al 2p in each layer of $MgB_2$ sample stack on sapphire as denoted by row labels. Mg KLL transitions are not assigned due to uncertainty in the composition and intensity of transitions in the observed chemistry. In the B cap layer, Mg KLL displayed 3 peaks, but likely originating from a Mg boride as no other elements were detected in this region. In the $MgB_2$ layer, an additional peak is found at a lower BE (higher KE, in Auger convention) which is likely from $MgB_2$. At the $MgB_2$/sapphire interface, an additional Mg KLL peak is observed at ~306 eV BE and the peak at 308 eV is intensified, indicative of a Mg oxide. No corresponding peak is observed in the B spectra so the interfacial species is likely primarily composed of Mg, Al and O. Once the interface is etched through, the primary signal is of sapphire ($Al_2O_3$). There is uncompensated shift in O 1s and Al 2p peaks due to differential charging of the sapphire substrate and $MgB_2$ overlayer, so binding energy of these peaks is not diagnostic of chemical states.**